\documentclass[bibyear]{aa} 
\usepackage{graphicx}
\usepackage{txfonts}
\usepackage{upgreek}
\usepackage{tipa} 
\usepackage{verbatim}
\usepackage{natbib}
\usepackage{color}

\begin{document} 

 \title{Vibrationally excited water emission at 658 GHz from evolved stars}

\titlerunning{Vibrationally excited water emission at 658 GHz from evolved stars}

   \subtitle{}

   \author{A. Baudry
          \inst{1} 
          \and
       E.M.L. Humphreys \inst{2} 
       \and
       F. Herpin\inst{1} 
       \and 
       K. Torstensson \inst{2} 
       \and 
       W.H.T. Vlemmings \inst{3} 
       \and 
       A.M.S. Richards \inst{4} 
       \and 
       M.D. Gray \inst{4} 
       \and 
       C. De Breuck  \inst{2} 
       \and M. Olberg \inst{3}
        }

   \institute{
   Laboratoire d'astrophysique de Bordeaux, Univ. Bordeaux, CNRS, B18N, all\'ee Geoffroy Saint-Hilaire, F-33615 Pessac, France 
              \email{alain.baudry@u-bordeaux.fr}
  \and
  European Southern Observatory (ESO), Karl-Schwarzschild-Str. 2, 85748 Garching bei Munchen, Germany 
  \and 
  Department of Space, Earth and Environment, Chalmers University of Technology, Onsala Space Observatory, S-439 92 Onsala, Sweden
  \and
  Jodrell Bank Centre for Astrophysics, School of Physics and  Astronomy, University of Manchester, Manchester M13 9PL, UK
  }
   
 \date{\today}   
 
 
  \abstract
   { Several rotational transitions of ortho- and para-water have been identified toward evolved stars in the ground vibrational state as well as in the first excited state of the bending mode  (v2=1 in (0,1,0) state). In the latter vibrational state of water, the 658 GHz $J = 1_{1,0} - 1_{0,1}$ rotational transition is often strong and seems to be widespread in late-type stars. 
      }
   {Our main goals are to better characterize the nature of the 658 GHz emission, compare the velocity extent of the 658 GHz emission with SiO maser emission to help locate the water layers and, more generally, investigate the physical conditions prevailing in the excited water layers of evolved stars. Another goal is to identify new 658 GHz emission sources and contribute in showing that this emission is widespread in evolved stars. 
      }
   {
   We have used the $J = 1_{1,0} - 1_{0,1}$ rotational transition of water in the (0,1,0) vibrational state nearly 2400 K above the ground-state to trace some of the physical conditions of evolved stars. Eleven evolved stars were extracted from our mini-catalog of existing and potential 658 GHz sources for observations with the Atacama Pathfinder EXperiment (APEX) telescope equipped with the SEPIA Band 9 receiver. The $^{13}$CO $J=6-5$ line at 661 GHz was placed in the same receiver sideband for simultaneous observation with the 658 GHz line of water. We have compared the ratio of these two lines to the same ratio derived from HIFI earlier observations to check for potential time variability in the 658 GHz line. 
 We have compared  the 658 GHz line properties with our H$_2$O radiative transfer models in stars and we have compared the velocity ranges of the 658 GHz and SiO $J=2-1$, v=1 maser lines. 
   }
   {Eleven stars have been extracted from our catalog of known or potential 658 GHz evolved stars. All of them show 658 GHz emission with a peak flux density in the range 
 $\approx 50-70$ Jy (RU Hya and RT Eri) to $\approx 2000-3000$ Jy (VY CMa and W Hya). Five Asymptotic Giant Branch (AGB) stars and one supergiant (AH Sco) are new detections. Three AGBs and one supergiant (VY CMa) exhibit relatively weak $^{13}$CO $J=6-5$ line emission while o Ceti shows stronger $^{13}$CO emission. We have shown that the 658 GHz line is masing and we found that the 658 GHz velocity extent tends to be correlated with that of the SiO maser suggesting that both emission lines are excited in circumstellar layers close to the central star. Broad and stable line profiles are observed at 658 GHz. This could indicate maser saturation although we have tentatively provided first information on time variability at 658 GHz.
 
 }

\titlerunning{Vibrationally excited water emission at 658 GHz from evolved stars}

   \maketitle
%

\section{Introduction}

Water is an abundant and widespread molecule of the Universe
with a rich spectrum of vibrational, rovibrational or pure rotational transitions spanning a broad range of wavelengths from the IR to the submm/mm or cm domains. Under non-local thermal equilibrium conditions, water exhibits moderate or very strong maser action in astrophysical sources as diverse as central regions of active galactic nuclei, star-forming regions, or envelopes of evolved stars (for evolved stars see e.g., the review on maser observations  and new model predictions by Gray et al. 2016). Strong 22 GHz emission from the $J = 6_{1,6} - 5_{2,3}$ transition of water in the ground vibrational state was first reported by \citet[][]{cheung1969} toward Orion. This transition has now been detected in hundreds of high-mass or low-mass young stellar objects (see e.g., \citet[][]{urquhart2011} or \citet[][]{furuya2001} and erratum in  \citet[][]{furuya2007}) and toward hundreds of evolved stars \citep[e.g.,][]{kim2014}). Maser action is strongly established at 22 GHz from Very Long Baseline Interferometry observations which show that the brightness temperature may reach 10{$^{12}$ or 
10{$^{12-15}$ K 
in evolved stars or star-forming regions for gas clouds as small as a few astronomical units.

In the ground vibrational state, maser emission has also been observed at frequencies higher than 22 GHz in the interstellar medium and toward evolved stars  for several transitions of ortho- and para-water (see e.g., Fig. 1 in \citet[][]{neufeld2017} for maser transitions known to date). Most of the observed line intensity ratios are rather well explained by collisional pumping (see model predictions of  \citet[][]{neufeld1991}, \citet[][]{yates1997}, \citet[][]{daniel2013} or \citet[][]{gray2016}). However,  in addition to collisional pumping, the  439 ($J = 6_{4,3} - 5_{5,0}$) and 471 ($J = 6_{4,2} - 5_{5,1}$) GHz transitions require radiation pumping from warm dust  (\citet[][]{yates1997} and \citet[][]{gray2016}). The 437 GHz ($J = 7_{53} - 6_{6,0}$) transition is another rather strong line in evolved stars (\citet[][]{melnick1993}, \citet[][]{menten2008}) but it is not  explained by collisional pumping alone or by the collisional and radiative pumping models of \citet[][]{gray2016} despite including recent collision rates and energy levels up to the second excited state of the bending mode (i.e., levels up to the (0,2,0) vibrational state). 
We also note that in the (0,2,0) state one strong rotational transition was identified toward VY CMa \citep[][]{tenenbaum2010} and one, or perhaps two, other weaker transitions were also observed toward IK Tau \citep[][]{velillaprieto2017}. Observing maser transitions in different vibrational states thus appears to be important not only to better constrain the current maser models and better explain lines in the ground and vibrationally excited states, but also to better probe the physical conditions and dynamics of a given region of a source or different regions in the same source. Of course, future H$_2$O modeling would also benefit from more accurate collisional excitation rates through the H$_2$O energy ladder.

Several  rotational transitions of water in the (0,1,0) state have been identified in evolved stars and most of them are probable masers (see review by \citet[][]{humphreys2007} and Tables 1 and 6 in \citet[][]{gray2016}). The first observations of relatively high $J$ rotational transitions in the first vibrational bending mode were made with the IRAM 30-m \citep[][]{menten1989} and the Atacama Pathfinder EXperiment (APEX)  telescope \citep[][]{menten2006}. In contrast with these weak lines,  the $J = 1_{1,0} - 1_{0,1}$ rotational transition of water at 658 GHz in the (0,1,0) vibrational state  may be strongly excited. The latter transition was first discovered by \citet[][]{menten1995} with the CSO 10-m telescope towards 10 out of 12 evolved stars and further observed by \citet[][]{hunter2007}, suggesting that the 658 GHz sources could be widespread. So far, the 658 GHz vibrationally excited line has only been detected in the envelopes of evolved stars with the exception of Orion source I mapped with ALMA \citep[][]{hirota2016}. 

The present work primarily aims at contributing to the characterization of the 658 GHz line emission observed in stars and at expanding the number of known 658 GHz sources. 
We attempt here to bring arguments showing that the 658 GHz line is masing and related to SiO maser line emission close to the central star. Because it is often strong, the 658 GHz line appears to be most useful to probe the dynamics and physical properties of the inner circumstellar gas layers. The targets observed in this work are  extracted from a larger sample of O-rich evolved stars that we have selected to show that the 658 GHz emission is widespread, hoping that some of them could later be mapped with ALMA. In light of the current model predictions for evolved stars that incorporate relevant collisional rates (\citet[][]{nesterenok2015} and \citet[][]{gray2016}) we try to determine the physical conditions prevailing in the 658 GHz-emitting clouds close to the central stars.

With these goals in mind we have observed the 658 GHz line toward a selected sample of Asymptotic Giant Branch (AGB) stars and Red Super Giant (RSG) stars during Science Verification of the SEPIA\footnote{Swedish-ESO PI receiver for APEX} Band 9 receiver installed on the APEX telescope (Sect. \ref{sec:observations}). Our spectral analysis, spectra, and line parameters are presented in Sect. \ref {sec:specresults}. The main characteristics of the 658 GHz emission and some of our modeling results are  discussed in Sect. \ref {sec:character}. Concluding remarks are given in Sect. \ref {sec:conclusion}.

\begin{table*}
\caption{Stellar sample and observation dates. }       
\label{source_list}      
\centering                          
\begin{tabular}{lccccccc}        
\hline                 
Star & Variability & RA & Dec  & Distance (error)$^{a}$ & Mass-loss rate $^{b}$ & Obs. date & Precipitable water 
\\   
                     & Spectral type & J2000 (h m s) & J2000 ($^{\circ}$ ' '') & (pc) &   (M$_{\odot}$yr$^{-1}$)   & (2016) & (mm)   \\
                      \hline
                      \hline
 o Ceti  & Mira & 02 19 20.79 & $-$ 02 58 39.5  & 92 (+11/$-$10) & 2.5 x 10$^{-7}$  & 26 July & 0.4 $-$ 0.5  
 \\
 & M5$-$M9 & & & & & \\
 
  R Hor $^{c}$ &  Mira & 02 53 52.72 &  $-$ 49 53 22.7 &  210 (+54/$-$36) & & 9 April  &  0.7 $-$ 0.8     
  \\
  & M5$-$M8 & & & & & \\
  
  RT Eri $^{c}$ & Mira &  03 34 12.48 &   $-$ 16 09 50.7  & 505 (+1800/$-$220) & & 15 June &  0.5 $-$ 0.65   
   \\ 
   & M7 & & & & & \\
   
 IK Tau & Mira &   03 53 28.87 &    +11 24 21.7  & 250$-$265 & 5 x 10$^{-6}$&   28 July & 0.65 
 \\ 
 & M6$-$M10 & & & & & \\
 
VY CMa & RSG &   07 22 58.32 &    $-$25 46 03.2  & 1200 (+130/$-$100) & 1.8 x 10$^{-4}$ &   15 June & 0.5$-$0.65 
 \\ 
 & M5 & & & & & \\
 
 RT Vir $^{c}$ & SRb &  13 02 37.98 &   +05 11 08.4  & 136 (+17/$-$14) & 1.3 x 10$^{-7}$ & 28 July &    0.8 
  \\  
  & M8 & & & & & \\
  
 R Hya & Mira & 13 29 42.78 &  $-$ 23 16 52.8 & 124 (+12/$-$10) & & 28 july & 0.8 
 \\
 & M6$-$M9 & & & & & \\
 
 W Hya & Mira/SRa & 13 49 02.00 & $-$ 28 22 03.05 & 98 (+30/$-$18) & 1.4 x 10$^{-7}$ & 28 July & 0.8 
 \\
 & M7$-$M9 & & & & & \\
 
  RU Hya $^{c}$ & Mira & 14 11 34.40 & $-$ 28 53 07.4 & --- & & 28 July & 0.75 
  \\
  & M6$-$M9 & & & & & \\
  
  AH Sco $^{c}$ & SRc/RSG & 17 11 17.02 & $-$ 32 19 30.7 & 2260 (+190/$-$190)& & 9 April & 0.45 $-$ 0.5 
  \\
  & M4$-$M5 & & & & & \\
  
  X Pav $^{c}$ & SRb & 20 11 45.86 & $-$ 59 56 12.8 & 671 (+2100/$-$382)& & 26 July & 0.4 $-$ 0.5 
  \\
  & M6$-$M7 & & & & & \\
  
\hline                                  
\end{tabular}
\tablefoot{$^{(a)}$Distances to o Ceti, R Hor, RT Eri, RT Vir, R Hya, and X Pav are from Hipparcos validated reduction \citep[][]{vanleeuwen2007}. The distance range for these stars is derived from the parallax uncertainty given in Hipparcos. The parallax uncertainty is large for RT Eri and X Pav, hence the large distance uncertainty. For RU Hya the uncertainty is larger than the estimated parallax and we do not give a distance entry. The distances to VY CMa \citep[][]{zhang2012}, W Hya \citep[][]{vlemmings2013} and  AH Sco \citep[][]{chenxi2008} are taken from VLBI observations. For IK Tau we give the range 250 to 265 pc from \citet[][]{olofsson1998} and \citet[][]{hale1997}, respectively. $^{(b)}$If known, mass-loss rates are taken from the literature. M$_{\odot}$yr$^{-1}$ is from \citet[][]{debeck2010} for VY CMa and W Hya, and \citet[][]{kimh} for IK Tau. $^{(c)}$New 658 GHz detection
 } 
\end{table*}


\section{Source sample and observations}
\label{sec:observations}
\subsection{658 GHz sources and new source selection}
\label{sec:sample} 

Following the discovery by \citet[][]{menten1995} of the (0,1,0) $J = 1_{1,0} - 1_{0,1}$ transition of water in ten stars, the low-spatial-resolution observations made with the SMA interferometer 
by \citet[][]{hunter2007}  expanded the 658 GHz source count to sixteen. Later, \citet[][]{justtanont2012} made a molecular inventory of nine O-rich AGB stars with HIFI aboard the Herschel satellite. The 658 GHz line of water was present in all stars. Herschel-HIFI also showed that there is weak 658 GHz emission from two OH/IR sources \citep[][]{justtanont2012} and one young proto-planetary nebula \citep[][]{bujarrabal2012}. As of today, and excluding the OH/IR sources, there are three supergiants and sixteen Mira or semi-regular variables known as 658 GHz emitters from the published literature. We also note that during the ALMA antenna commissioning period, two 7-m antennas were used at the 2900-m high site (OSF) to survey several stars in the 658 GHz line of water and detect known or new emitting sources \citep[][]{phillips2017}. 

Observing the 658 GHz line is difficult because of low atmospheric transparency and relatively high receiver noise temperature in Band 9. However, as argued earlier in the introduction, surveying  a large number of selected O-rich stars with APEX should allow us to discover new 658 GHz sources. The brightest ones could later be mapped with ALMA to unravel the spatial structure of the 658 GHz emission and probe the inner layers of the circumstellar envelope. 
With these goals in mind we have built a mini-catalog of nearly 100 potential and known 658 GHz sources visible from the southern sky based on stars with known SiO (43/86 GHz) and H$_{2}$O (22 GHz) maser emission above a fixed flux density limit. Most of these 100 sources were extracted from the homogeneous sample of evolved stars observed simultaneously in SiO and H$_{2}$O (22 GHz) by \citet[][]{kim2010} and \citet[][]{cho2012} after we had restricted our selection to flux density around or larger than about 50 Jy. To improve the negative declination source coverage we have incorporated into our initial catalog additional sources meeting the same flux density criterion and taken from other SiO publications \citep[e.g.,][]{haikala1994}. We have also used the compilation of \citet[][]{benson1990} to help identify the SiO/H$_{2}$O sources that can be observed from the APEX site. 

For these first 658 GHz observations made with APEX,  
we have extracted nine Mira-type or semi-regular variables and two supergiants from our mini-catalog (see Table \ref{source_list}) in order to  detect new emission sources and/or confirm earlier results without published spectra. 


\subsection{Observations}
\label{sec:observsetup}
We have used APEX SEPIA Band 9 Science Verification time under program code ESO 097.F-9806A to perform observations of the nine AGB stars and the supergiant AH Sco shown in Table \ref{source_list}. VY CMa data were acquired separately (see below). SEPIA Band 9 receiver is a dual polarization, double sideband (DSB) receiver\footnote{Band 9 DSB receiver will soon be upgraded to a sideband-separating receiver}  with technical specifications similar to those of the ALMA Band 9 receiver  \citep[][]{baryshev2015}. It was tuned to place the $J = 1_{1,0} - 1_{0,1}$ transition of water at 658.007 GHz and the $J = 6-5$ line of $^{13}$CO at 661.067 GHz in the receiver lower sideband (LSB). The receiver LSB and upper sideband (USB) frequency ranges were 657.537 to 661.537 GHz and 669.537 to 673.537 GHz, respectively. 
 
 The VY CMa data were acquired on 15  June, 2016  (Table \ref{source_list}), with the same DSB receiver on APEX under program code 296.D-5052 (A). Two different tunings were used to detect in the LSB $^{28}$SiO v=0 $J = 15-14$ at 650.958 GHz (tuning 1) and $^{29}$SiO v=0 $J = 15-14$ at 642.807 GHz and $^{28}$SiO v=1 $J = 15-14$ at 646.431 GHz (tuning 2). The image sideband (USB) included the 658.007 GHz  line of water (tuning 2) and the  $^{13}$CO at 661.067 GHz (tuning 1). W Hya was also observed on the same day and comparison of the 658 GHz results with data acquired on 28 July, 2016, is briefly discussed in  Sect. \ref{sec:variability}. 

According to the `splatalogue' database the exact frequency of the $J = 1_{1,0} - 1_{0,1}$ transition is 658.00655 GHz or 658.00625 GHz from another determination; this implies a frequency indetermination of 0.3 MHz or 0.14 km~s$^{-1}$. Telescope pointing was generally done on the source by observing the $J=6-5$ line of CO and the telescope focus was adjusted on a planet. For R Hor, however, no pointing was possible but a system check was performed by CO  $J=6-5$ line detection. The half-power beam width (HPBW) at 658 GHz is $ \approx 9 \arcsec$. Because of this narrow beam we also checked for potential pointing drifts (see Sect. \ref {sec:results}) although the targets were observed during relatively short integration times ranging from 15 to 35 minutes. 
The observations were made with good weather conditions in position-switching mode. The precipitable water vapor varied from 0.4 to 0.8 mm during our observations (see Table \ref{source_list}). Flux density calibration is described in Sect. \ref {sec:calib}. 

To achieve 4 GHz coverage 
we overlapped the two APEX FFT spectrometers, each spectrometer providing 2.5 GHz instantaneous bandwidth.

\subsection{Flux density calibration}
\label{sec:calib}

Accurate flux density calibration is difficult at the high 658 GHz frequency of the Band 9 receiver. We have used two different approaches to estimate the flux density to antenna temperature conversion factor, $S/{T_A}^\star$, where $S$ is the flux density and ${T_A}^\star$ the antenna temperature for outside the atmosphere.

We first observed planets and especially Mars at the time of our 2016 observations. These observations give $S/{T_A}^\star$ = 120 ($+/-$15) Jy/K. In the second approach, we used the $^{13}$CO $J=6-5$ line at 661 GHz to compare the sources detected in this work with the independent HIFI observations of the same sources \citep[][]{justtanont2012}. (We did not use the stronger 658 GHz line of water for this comparison because it is masing and could be affected by time-variability; see Sect. \ref {sec:variability}). The HIFI data are calibrated in main beam temperature ($T_{mb}$) and we have used a beam width of 32\arcsec at 661 GHz to transform the velocity integrated $T_{mb}$ given in \citet[][]{justtanont2012} into an integrated flux density which, divided by our integrated antenna temperature (see Table \ref {table_param13CO} ), gives $S/{T_A}^\star$. Our strongest detection and best $^{13}$CO $J=6-5$ data are obtained for o Ceti from which  we derive $S/{T_A}^\star$ = 185 Jy/K with an uncertainty around 30 Jy/K (mainly due to the HIFI intensity scale uncertainty). Another independent line approach to estimate the conversion factor is based on the relation between the main beam temperatures obtained with the CHAMP+ heterodyne receiver on APEX and the peak antenna temperatures observed with the SEPIA Band 9 receiver for different AGB stars. We obtain a conversion factor above 185 Jy/K but the uncertainty in the $T_{mb}$ versus  ${T_A}^\star$ slope depends on the number of stars used. 

In conclusion, our independent approaches give 120 ($+/-$15) or 185 ($+/-$30) which implies a rather large uncertainty in the $S/{T_A}^\star$ conversion factor. This result is not fully understood but we attribute the high Jy/K conversion factors derived here to the `pick-up' mirror placed in front of the input window of the Band 9 receiver (this mirror which degraded  the beam coupling was later 
replaced by a new one).

\section{Spectral analysis and results}
\label{sec:specresults}

\subsection{Spectral analysis}
\label{sec:spectral}

The Observatory delivered on-line standard opacity-calibrated data using one atmospheric mean opacity across 2.5 GHz. We checked that off-line recalibration of our 658 GHz data (using opacity values across several channels) gave consistent results with the spectra obtained with the on-line standard calibration. Therefore, we do not use here off-line recalibrated data. 

Further spectral analysis was performed with the CLASS package\footnote{http://www.iram.fr/IRAMFR/GILDAS/}. All subscans were first examined with maximum spectral resolution (around 0.015 km~s$^{-1}$ in the LSB) to search for eventual spurious signals in the signal band (LSB) or signals leaking from the USB into the signal band. The individual scans were averaged as well as the two polarization signals and, after baseline subtraction, all spectra were binned to a lower resolution. We used a resolution of 0.14 km~s$^{-1}$ for the 658 GHz H$_{2}$O  spectra, in agreement with the H$_{2}$O rest frequency indetermination mentioned in Sect. \ref {sec:observsetup}. We have also verified that for some sources spectral details could be lost at 658 GHz with lower resolution (e.g., $0.5-1$ km~s$^{-1}$). For the much broader $^{13}$CO $ J = 6-5$ line, our spectra were binned to 1.1 km~s$^{-1}$, except for VY CMa where we used 3 km~s$^{-1}$. A first order baseline subtraction was used to derive the line parameters given in  Tables \ref{table_param} and \ref{table_param13CO}.

\subsection{Results}
\label{sec:results}

The H$_{2}$O line parameters are gathered in Table \ref{table_param} and the spectra  for the AGB stars and two supergiants in our sample are displayed in Figs. \ref{658spec} and \ref{supergiants} with 0.14 km~s$^{-1}$ resolution (0.17 km~s$^{-1}$ for VY CMa). The first three columns in Table \ref{table_param} give the source name, the velocity at peak intensity and peak intensity in terms of antenna temperature for outside the atmosphere, ${T_A}^\star$. Relatively strong or weak 658 GHz emission is observed for stars in our sample with a maximum of 17.3 K for W Hya, while  for RU Hya and RT Eri we observe ${T_A}^{*}_{peak}$ $\approx 0.3-$0.4 K. The  corresponding flux density range is $\approx$ 50 Jy to 2000 or 3000 Jy (depending on the adopted conversion factor, see Sect. \ref {sec:calib}). This  is similar to the flux density range in \citet[][]{menten2006}.

The 658 GHz line profiles are smooth and tend to be asymmetrical but with no individual spectral components as narrow as those observed at 22 GHz in the same sources. 
For comparison with other thermally or quasi-thermally excited lines, we give a measure of the width of the 658 GHz line in terms of velocity at half peak intensity. This measure is independent of the line shape which will be discussed in Sect. \ref {sec:character}. The fifth column in Table \ref{table_param} gives the r.m.s. spectral noise achieved in one 0.14 km~s$^{-1}$ channel (in one 0.17 km~s$^{-1}$ channel for VY CMa) outside the spectral features. The last column gives the velocity integrated antenna temperature as directly determined from the calibrated antenna temperatures for all individual channels above the three sigma level.

The $^{13}$CO $J=6-5$ line is weakly detected in four sources, including VY CMa, and is relatively strong in o Ceti (see Fig. \ref{13COspec}). The velocity span observed here for the $J = 6-5$ line is consistent with the data reported in other works for lower $J$ levels of CO or $^{13}$CO \citep[e.g.,][]{debeck2010}. All line parameters in Table \ref{table_param13CO} are derived from Gauss-fits to our spectra. In o Ceti two components best match our data and they are both given in Table \ref{table_param13CO}.   
We note that the $^{13}$CO $J=6-5$ line was also weakly detected in three out of the ten 658 GHz sources observed by \citet[][]{menten1995}. 

Our line parameters in Tables \ref{table_param} and  \ref{table_param13CO} could mainly be affected by $\it (i)$ possible telescope pointing problems, or $\it (ii)$ line contamination of the signal band from the other sideband.  

$\it (i)$ We have frequently checked the telescope pointing (except for R Hor, see Sect. \ref{sec:observsetup}) and checked the focus during our observations. However, in the case of W Hya, plotting the H$_{2}$O line integrated intensity of each subscan versus time showed that the intensity had decreased during the observation. We had indeed to apply a pointing correction to the end of our observation and we have interpreted the intensity decrease as being due to rapid ambient temperature changes during the observing run. W Hya was also observed on the same day but with different spectral settings and we conclude that its `absolute' peak intensity cannot be estimated to better than about 15$\%$. Checks of integrated intensity versus time 
were also made for other strong sources and did not show significant variations from subscan to subscan, except for VY CMa. In the latter source, only the data not affected by a pointing drift were kept (from which we derived the parameters shown in Table \ref{table_param}).  

$\it (ii)$ Line contamination of the signal band from molecular lines falling in the other receiver band cannot be excluded a priori. For the ESO 097.F-9806A project we have used  the `splatalogue' database to check if lines that could possibly be excited in the envelope of evolved stars fall in the 669. 53 to 673.54 GHz  range (receiver USB). This included CO and isotopes, SO and isotopes, SO$_2$, CS, HCN, SiS v=0, $^{29}$SiO, SiO v= 0$-$4 and  ground- and vibrationally excited-states of H$_2$O. We found that one line of SO and SiS and two SO$_2$ lines with relatively low energy (below about 
350 K) 
fall in the USB. Only the $J = 9_{9,1} - 10_{8,2}$ transition of SO$_2$ at  673.06713 GHz lies close to the 658.0063 GHz of water after the USB is folded into the signal band. This corresponds to a velocity difference of about 33 km~s$^{-1}$ which is outside the 658 GHz line profile according to the observed velocity range. Several SO$_2$  lines have been detected towards AGB stars \citep[][]{danilovich2016} but the emission is weak and the SO$_2$ $J = 9_{9,1} - 10_{8,2}$ transition was not observed by Danilovich et al. (although higher-energy transitions were observed). Therefore, we can reasonably conclude that our 658 GHz line observations made with the Band 9 DSB receiver are not significantly affected by line contamination.  
        
Line contamination of the broad $^{13}$CO $J=6-5$ line profiles shown in Fig. \ref{13COspec} was not investigated in depth. We simply note here that the velocity span shown in Fig. \ref{13COspec} coincides with that obtained with other instruments in the same sources for various rotational transitions of CO and $^{13}$CO \citep[e.g.,][]{debeck2010}.

  \begin{figure}
   \centering
   \includegraphics[scale=0.52, angle=0]{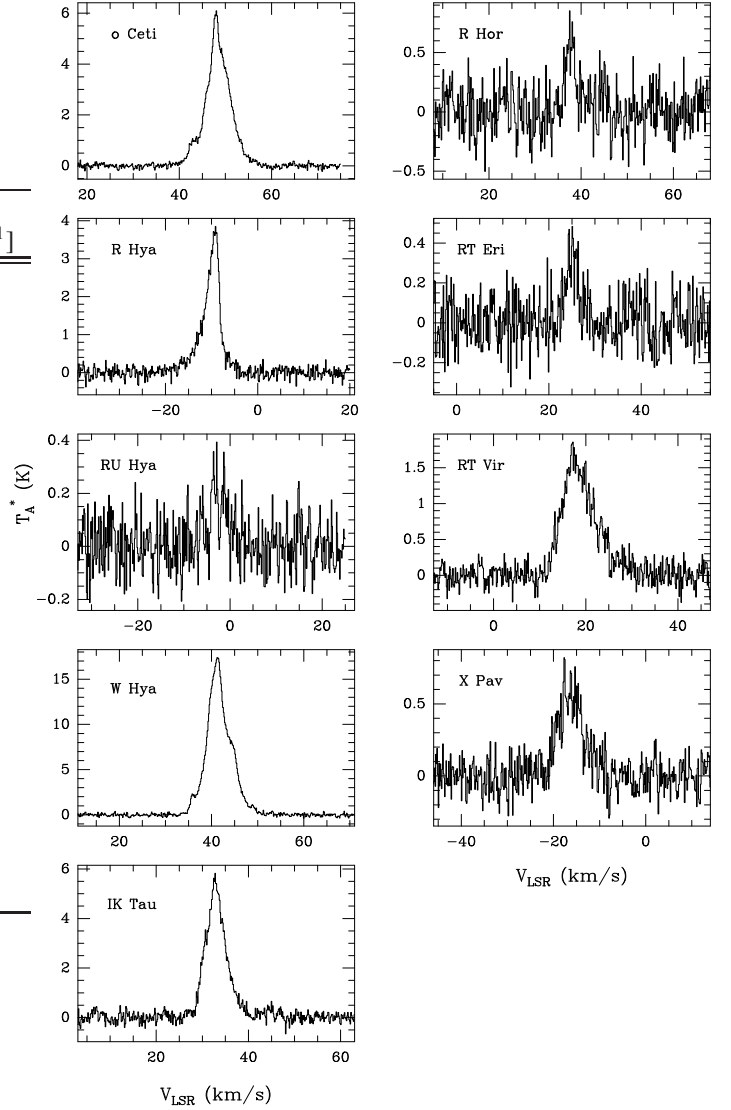}
   \caption{Spectra of the $(0,1,0), J = 1_{1,0} - 1_{0,1}$ transition of water at 658 GHz toward nine Mira-type or semi-regular variables in our sample. The antenna temperature is for outside the atmosphere and the spectral resolution is 0.14 km~s$^{-1}$ for all sources.  See Sect. \ref {sec:calib} for Jy to K conversion at the time of the observations.
   }
              \label{658spec}%
    \end{figure}
  
    \begin{figure}
   \centering
   \includegraphics[scale=0.46, angle=0]{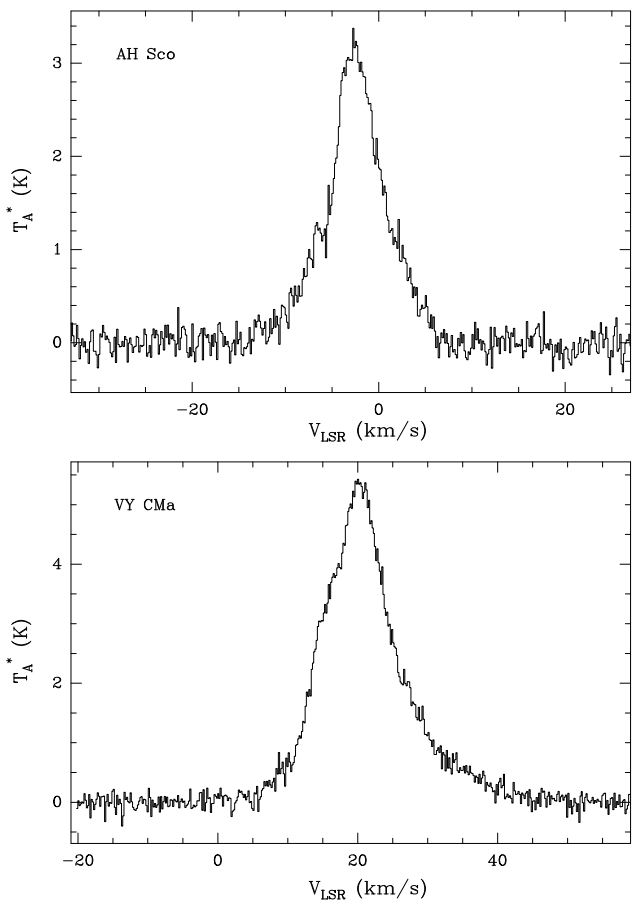}
   \caption{Spectra of the $(0,1,0), J = 1_{1,0} - 1_{0,1}$ transition of water at 658 GHz toward the red supergiants AH Sco and VY CMa, plotted at 0.14 and 0.17 km~s$^{-1}$ resolution, respectively. The antenna temperature is for outside the atmosphere. See Sect. \ref {sec:calib} for Jy to K conversion at the time of the observations.
   }
              \label{supergiants}%
    \end{figure}

    \begin{figure}
   \centering
   \includegraphics[scale=0.67, angle=0]{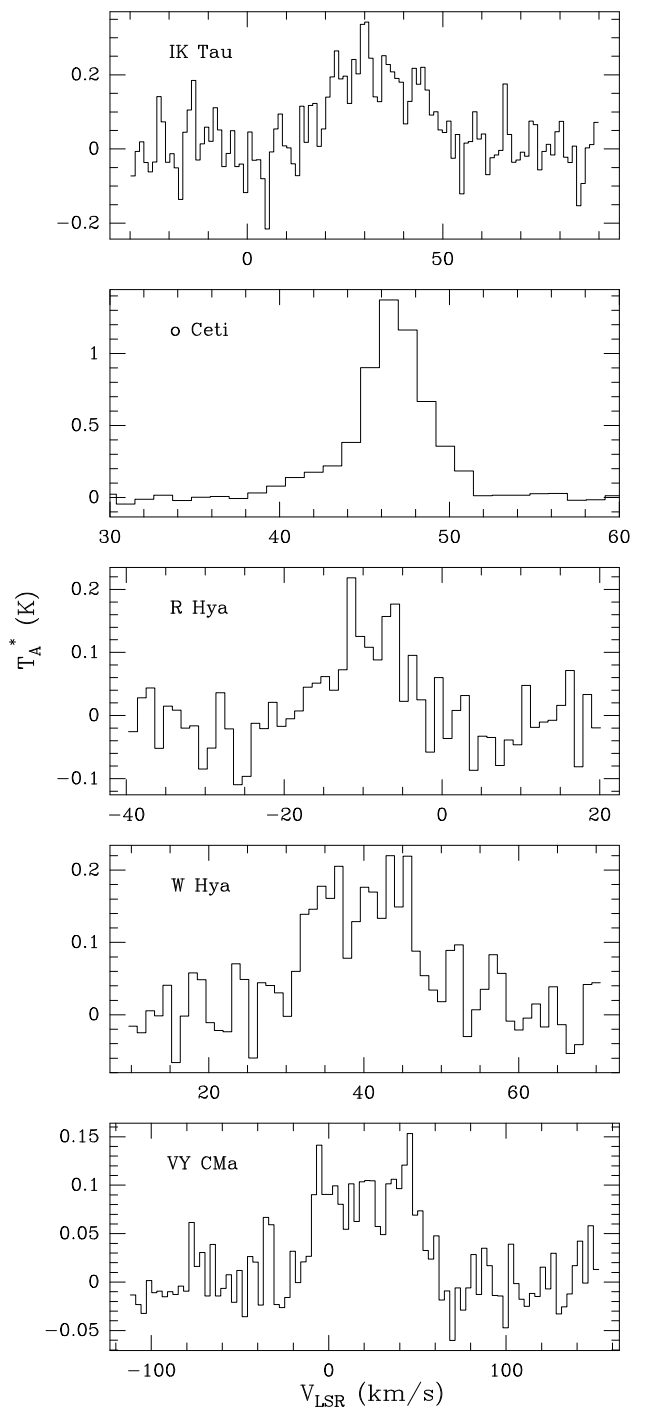}
   \caption{Spectra of the  $J=6-5$ transition of $^{13}$CO at 661 GHz toward five stars in our sample, plotted at 1.1 km~s$^{-1}$ resolution (3.0 km~s$^{-1}$ for VY CMa). The antenna temperature is for outside the atmosphere.  See Sect. \ref {sec:calib} for Jy to K conversion at the time of the observations.
   }
              \label{13COspec}
    \end{figure}


\begin{table}
 \caption{Observed H$_2$O line parameters. $\varv_{peak}$ and ${T_A}^{*}_{peak}$ are the velocity and antenna temperature at the 658 GHz line peak. $\Delta\varv_{1/2}$ is the  full line width at half-maximum for comparison with thermally excited lines. The r.m.s. in the fifth column is the noise outside the 658 GHz features for one 0.14 km~s$^{-1}$ channel (0.17 km~s$^{-1}$ for VY CMa).   In the last column we give the line area, $\int {{T_A}^{*}}$dv, for all channels three times above the r.m.s.
 }
\label{table_param}      
\centering
\begin{tabular}{lccccc} \hline 
Source &$\varv_{peak}$ & ${T_A}^{*}_{peak}$ & $\Delta\varv_{1/2}$ & r.m.s. & $\int {{T_A}^{*}}$dv  \\ 
& [km~s$^{-1}$] & [K] & [km~s$^{-1}$]  & [K] &  [K.km~s$^{-1}$]  \\ 
\hline   
\hline                     
 o Ceti $^{b}$ &  48.1 & 6.1  & 4.8 & 0.08 & 32.5 \\ 
 & & & & &  \\
 
 R Hor $^{a}$  &  37.7 & 0.6  & 2.4 & 0.19 & 1.0 \\ 
 & & &  & &  \\
 
 RT Eri $^{a}$  &  24.9 & 0.4  & 3.3 & 0.11 & 0.7 \\ 
 & & & & &  \\
 
 IK Tau $^{b}$ &  32.8 & 5.8  & 4.4 & 0.19 & 27.2 \\ 
 & & & & &  \\
 
 VY CMa $^{d}$ & 20.1 & 5.3 & 10.7 & 0.12 & 51.0 \\
 & & & & &   \\
 
RT Vir $^{a}$  &  17.0 & 1.8  & 7.6 & 0.14 & 12.4 \\ 
& & & & &  \\

 R Hya $^{c}$  &  $-$9.1 & 3.8  & 2.9 & 0.13 & 14.2 \\ 
 & & & & &  \\
 
  W Hya $^{d}$  &  41.3 & 17.3  & 4.4 & 0.14 & 97.2 \\ 
  & & & & &  \\
  
 RU Hya $^{a}$  &  $-$2.8 & 0.3  & 4.0 & 0.09 & 0.5 \\ 
 & & & & &  \\
 
AH Sco $^{a}$  &  $-$2.7 & 3.2  & 5.8 & 0.12 & 22.5 \\ 
& & & & &  \\

X Pav $^{a}$ &  $-$16.4 & 0.6  & 6.3 & 0.11& 3.2 \\ 
& & & & &  \\
   
\hline
\end{tabular}
\tablefoot{$^{(a)}$New 658 GHz detection, this work. $^{(b)}$First detected with HIFI. $^{(c)}$\citet[][]{hunter2007}, no spectrum shown. $^{(d)}$First detected by \citet[][]{menten1995}
 }    
\end{table}


\begin{table}
 \caption{$^{13}$CO $J=6-5$ line parameters in detected sources. $\varv_{peak}$, ${T_A}^{*}_{peak}$, $\Delta\varv_{1/2}$ and $\int {{T_A}^{*}}$dv are obtained from Gauss-fits to the spectra in  Fig. \ref {13COspec} for 1.1 km~s$^{-1}$ resolution (3.0 km~s$^{-1}$ in VY CMa). The r.m.s. uncertainties are given in parenthesis. 
}
\label{table_param13CO}      
\centering
\begin{tabular}{lcccc} \hline 
Source &$\varv_{peak}$ & ${T_A}^{*}_{peak}$ & $\Delta\varv_{1/2}$ & $\int {{T_A}^{*}}$dv  \\    
& (r.m.s.) & (r.m.s.) & (r.m.s.) & (r.m.s.) \\
& [km~s$^{-1}$] & [K] & [km~s$^{-1}$]  & [K.km~s$^{-1}$]  \\ 
\hline   
\hline                     
  o Ceti  &  46.7 & 1.42  & 3.6 & 5.38 \\ 
  & (0.0) & (0.03) & (0.1) & (0.12) \\
  &  50.2 & 0.20  & 2.0 & 0.42 \\ 
  & (0.2) & (0.03) & (0.3) & (0.09) \\
\\
  IK Tau  &  31.9 & 0.18  & 23.5 & 4.47 \\ 
  & (1.4) & (0.05) & (2.6) & (0.51) \\
  \\
  VY CMa  &  21.3 & 0.11  & 59.0 & 6.74 \\ 
  & (2.1) & (0.03) & (4.2) & (0.46) \\
  \\
 R Hya  &  $-$9.5 & 0.17  & 11.6 & 2.16 \\ 
 & (0.6) & (0.04) & (1.3)  & (0.21) \\
 \\
 W Hya  &  40.1 & 0.18  & 16.5 & 3.10 \\ 
 & (0.7) & (0.04) & (1.6) & (0.25) \\
  
\hline
\end{tabular}
\end{table}

\section{Characterization of the 658 GHz line emission} 
\label{sec:character}
\subsection{On the nature of the 658 GHz emission}
\label{sec:nature_658}

The vibrationally excited line profiles of water at 658 GHz do not show the usual narrow features or spectral complexity observed for strong water masers at 22 GHz. However, 
compared to the widths of the CO or $^{13}$CO lines in the same sources, the 658 GHz linewidths, as measured at half peak intensity, are relatively narrow. This is expected since the CO line provides an estimate of the expansion of the circumstellar envelope and is excited well beyond the few stellar radii where we think the 658 GHz line, with  energy levels around 2360 K, is formed. For gas temperatures in the range 1500 to 2500 K we expect thermal linewidths of order 1.9 to 2.5 km~s$^{-1}$ while, independently of the actual physical processes leading to the observed 658 GHz line shape, the 658 GHz half intensity widths are broader than these thermal linewidths. As further discussed in Sect. \ref {sec:variability}, the observed 658 GHz profiles are probably broadened by spatial blending of individually narrower components within our single dish beam. This was in fact demonstrated with ALMA for VY CMa \citep[][]{richards2014} and could explain why the 658 GHz line profiles are commonly asymmetric with single antennas. 

Other considerations indicate that the 658 GHz transition is not thermally excited. The flux density of this transition can reach several thousand Jy and we know that the line brightness temperature, ${T_b}$, is well above the gas kinetic temperature despite ${T_b}$ 
is not well constrained from single dish observations. In VY CMa, \citet[][]{richards2014} mapped the 658 GHz emission and 
identified spatial components with ${T_b} \approx 0.3-4$ 10{$^{7}$} K. Such temperatures indicate maser emission and we note that these values should be considered as lower limits because the true maser components are probably less extended than the ALMA beam used by \citet[][]{richards2014}. 
We can also use the nearly contemporaneous 22 GHz maser observations made by \citet[][]{menten1995}  to tentatively constrain the brightness temperature at 658 GHz. Their published data indicate a flux density ratio S(658)/S(22) in the approximate range $1-20$ for several 658 and 22 GHz features. Assuming that both emissions have similar spatial extents, we derive  ${T_b(658)} = {T_b(22)}$ x $S$(658)/$S$(22) x (1.14  10{$^{-3}$). The 22 GHz brightness temperature reaches 10{$^{7-12}$ K in AGBs and supergiants (see e.g., \citet[][]{richards2011}), so we expect ${T_b(658)}$ in the range 10{$^{4-10}$ K which clearly indicates suprathermal emission at 658 GHz.

Further indication that the 658 GHz line is masing is provided from radiative transfer calculations applied to typical conditions met in evolved stars, see  \citet[][]{gray2016} and \citet[][]{nesterenok2015}. \citet[][]{gray2016} have considered material slabs in the range 4.5 10{$^{13}$  to 2.25 10{$^{14}$ cm to show that the 658 GHz transition can easily be inverted. Conditions giving rise to 658 and 22 GHz maser emission are compared in Fig. \ref{fig22_and_658} where we have plotted negative maser depth contours  in the H$_2$ density/kinetic temperature plane. Negative opacities may reach 10 or more at 658 GHz for kinetic temperatures ($\approx800-$2800 K) higher than those at 22 GHz  and for relatively high densities (above 10{$^{10}$ cm{$^{-3}$) suggesting material layers close to the stellar photosphere.  
Model predictions indicate that the 658/22 opacity ratios can reach about 5 in evolved stars for regions with a moderate dust temperature of order 50 (see Table 6 in Gray et al. 2016). 
However, such a ratio cannot be used to estimate  ${T_b(658)}$ from  ${T_b(22)}$ since it just reflects the ratio of the maximum maser opacity at 658 GHz to the maximum obtained at 22 GHz without accounting for saturation effects and differing radiation field conditions. As expected, Fig. \ref{fig22_and_658} shows that at 22 GHz the loci of inverted regions are broader than at 658 GHz for a transition where both collisions and radiative pumping contribute to 22 GHz maser excitation.

An independent study made by  \citet[][]{nesterenok2015} also shows that the 658 GHz transition can be strongly inverted. His modeling provides for example an amplification gain of about  10{$^{-13}$ cm{$^{-1}$ for a H$_2$ density of 5 10{$^{9}$ cm{$^{-3}$ and an o-H$_2$O to H$_2$ ratio of  6 10{$^{-5}$. This gain is achieved for kinetic temperatures above 1000$-$1200 K and corresponds to an opacity of several units in a cloud of several astronomical units. As in \citet[][]{gray2016}, the 658 GHz maser is found to be collisionally pumped and shows a strong dependence on kinetic temperature.

\begin{figure}
  \centering
   \includegraphics[bb = 07 100 532 879, clip, width=7.8cm, angle=270]{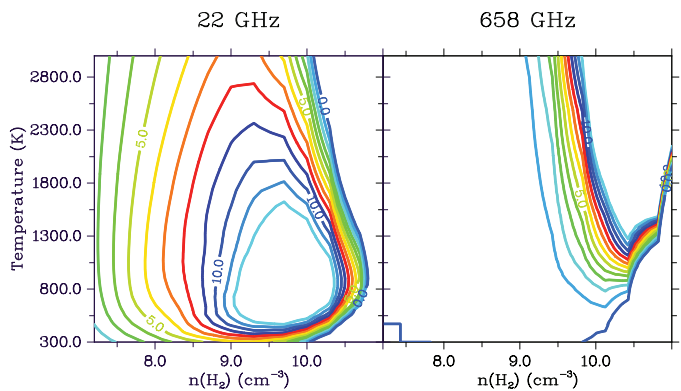}
   \caption{Maser negative optical depths at 22 and 658 GHz in the kinetic temperature and H$_2$ density plane. The contours are in regularly spaced intervals of $\delta$(optical depth) = 1. The highest contour is for an optical depth of 12 at both 22 and 658 GHz.}
     \label{fig22_and_658}
    \end{figure}

\subsection{Where is the 658 GHz line excited?}
\label{sec:localization_658}

It is not well known in which atmospheric or circumstellar layers the 658 GHz line is formed, except for VY CMa which was mapped with ALMA \citep[][]{richards2014}. The ALMA map shows an aggregation of many individually unresolved emission spots within roughly 50$-$250 mas of the central star. The overall emission structure is complex and shows organized spatial features within or close to the SiO maser spatial extent \citep[][]{richter2013} but also more distant structures, perhaps related to shocks.

To indirectly check if the 658 GHz water emission is excited close to the central star, we have compared the 658 GHz velocity extent with the same quantity for the SiO maser emission in the first vibrational state. This is justified because: $\it(i)$ the SiO, v=1 energy, $\approx 1780 K$, 
is close to 2300 K
 for the (0,1,0) vibrational state of water at 658 GHz; $\it(ii)$ the emission peak velocities of both species are always close to each other; and $\it(iii)$ the SiO masers are excited within about 5 stellar radii as shown from VLBI observations. To this end, we have compared for ten objects in the present 658 GHz source sample (we have excluded our weakest object RU Hya), the full velocity width at zero intensity, FWZI(658), with the same quantity for the SiO, v=1, $J=2-1$ line at 86.243 GHz, FWZI(86). The 86 GHz velocity width is taken at the 2 sigma level from our catalog of polarized SiO maser emission in evolved stars \citep[][]{herpinbaudry2003}. The latter work includes line parameters for 120 objects observed in 1999 and 2000 with the IRAM 30-m telescope. To expand this comparison to more sources we have estimated FWZI(658) from the published 658 GHz spectra for seven and three other objects taken from \citet[][]{menten1995} and \citet[][]{hunter2007}, respectively. We are aware that determining FWZI(658) from the literature without access to the original data is uncertain (these data are shown as blue dots in Fig. \ref{fig_vel_correl}) but we have a good determination of FWZI(86) from our IRAM SiO catalog for eight of these ten additional sources. For the three sources not observed from the IRAM site (R Hor, RT Eri and X Pav) we have used the 86 GHz SiO spectra of \citet[][]{haikala1994} and  \citet[][]{haikala1990} to estimate FWZI(86). There is another complication because our 30-m data have shown that FWZI(86) can vary with the epoch of the observation. This is the case for instance of TX Cam where the SiO activity may change the value of FWZI(86) by a few  km~s$^{-1}$.

Despite these uncertainties and difficulties, Fig. \ref{fig_vel_correl} suggests that FWZI(658) tends to be correlated with FWZI(86). This trend extends to high velocities for three of the supergiants in this plot, VX Sgr, NML Cyg and, especially, VY CMa. AH Sco does not show an exceptional velocity extent at both 658 and 86 GHz. The trend observed here suggests 
that the bulk of the 658 GHz and 86 GHz emission lines are excited in the same circumstellar material or that similar conditions are required to excite the 658 and 86 GHz lines. It should be confirmed, however, with a larger source sample and a more homogeneous data set.

\subsection{Comments on 658 GHz relative intensities and polarization}
\label{sec:relative_intensity}

Our relative peak intensities (see Table \ref{table_param}) are rather large and range from $\approx$ 40 : 1 or 60 : 1 when one compares W Hya with RT Eri or RU Hya. This is similar to $\approx 30:1$ derived from \citet[][]{menten1995} by comparing their strongest source, W Hya, with their weakest detection, S CrB.   
A large range of relative intensities could be related to various factors such as the relative distance of the objects and the exact nature of the maser (degree of saturation for instance). We first note that in the case of RT Eri or RU Hya the distance is highly uncertain or unknown because the parallax uncertainty from Hipparcos is nearly as large as the parallax itself or larger. The same is true for X Pav for which the distance could be anything between $\approx 300$ and $2800$ pc. For these three stars the distance argument could perhaps explain part of the the observed weak emission in comparison with W Hya which is only $\approx 100$ pc away from us. Other factors related to the details of the maser pumping scheme and the degree of saturation could play a role in the observed relative intensities. Predictions would require a model for each individual source and we also note that comparison of two fully saturated masers (although saturation cannot be demonstrated from our 658 GHz data, see Sect. \ref {sec:variability}) would vary in proportion with the maser opacity ratio. The pumping conditions may also vary with the optical phase of the central star (e.g., density variations implying different collision rates) and the maser luminosity might depend on different pumping rates.

The polarization properties of the 658 GHz maser emission in stars are unknown; these should be considered when comparing different stars or the same object at different epochs. Stellar masers are often linearly polarized and some features also show weak circular polarization. 
In the present work, a clean separation of the two linear polarizations is nearly impossible because the level of leakage of each polarization into the other is unknown. In addition, our parallactic angle coverage was small with rather short integration times. Dedicated observing sessions using the future Band 9 sideband separating receiver are needed to extract valuable polarization information at 658 GHz.  

 \begin{figure}
  \centering
  \includegraphics[scale=0.35, angle=0]{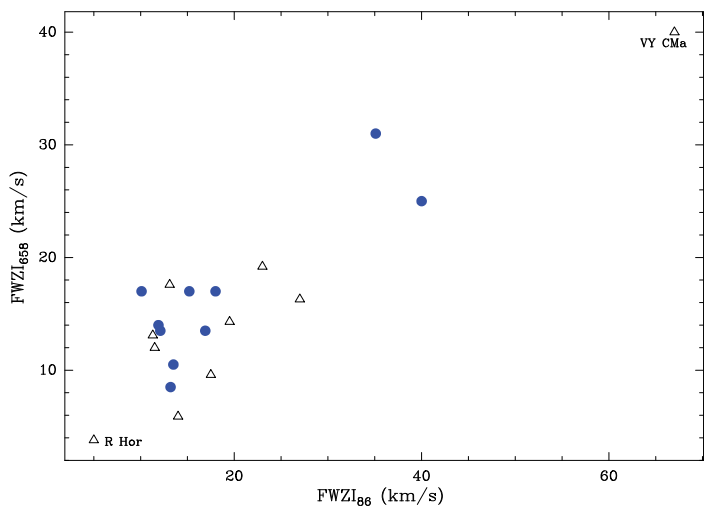}
   \caption{Full width at zero intensity (FWZI) at 658 GHz (H$_2$O, (0,1,0) vibrational state) versus FWZI at 86 GHz (SiO, v=1) for ten sources observed in this work (triangles); R Hor and VY CMa are labeled at the two ends of this plot, and RU Hya is not included because of very weak emission. Ten other additional sources discussed in the text are shown as blue dots.
}
   \label{fig_vel_correl}
  \end{figure}

\subsection{658 GHz line profile stability and comments on time variability}
\label{sec:variability}

The 658 GHz line profiles obtained by different authors for two sources in our sample, VY CMa and W Hya, exhibit a great stability over a long period of time. For VY CMa, in addition to our APEX spectra of 2016, there are published spectra in \citet[][]{menten1995}, \citet[][]{hunter2007}, \citet[][]{alcolea2013}, and \citet[][]{richards2014} covering a period of 21 years. Different channel spacings were used in these works but all show the same asymmetrical profile with a shoulder around 12$-$15 km~s$^{-1}$ and a broad, weak pedestal emission. On the other hand, the peak intensity as reported from these observations tends to show a regular decline with time (see end of this Section).
For W Hya, the spectra published by \citet[][]{menten1995} and \citet[][]{hunter2007} exhibit the same main broad feature and shoulder-like feature observed by us around 41.3 km~s$^{-1}$ and  44$-$45 km~s$^{-1}$, respectively. Only our data from June and July 2016, acquired with higher spectral resolution than the 0.4 and 0.9 km~s$^{-1}$ in \citet[][]{hunter2007} and  \citet[][]{menten1995}, show an additional weak feature at 35.9 km~s$^{-1}$. These observational results indicate that there is long-term stability of the overall line profile at 658 GHz and no short-term variability on one-month scales since both the main peak and the weaker feature are present in our 2016 data. We cannot exclude some time variability at 658 GHz, however, because of the newly detected weak 35.9 km~s$^{-1}$ feature in our data. The origin of the 15$\%$ peak intensity variation observed in W Hya between our June and July 2016 data is uncertain. This relatively small intensity change observed for two different frequency settings on different days seems consistent with our calibration uncertainties. 

The linewidths observed at 658 GHz (Table  \ref {table_param}) are much broader than the thermal linewidths ($\approx2-$2.5 km~s$^{-1}$) and the total velocity extent most probably results from the aggregation of multiple 658 GHz maser components that we do not spatially separate within the APEX beam. The ALMA observations toward VY CMa have shown indeed multiple spatial maser components associated with different velocity components across the broad single dish spectrum. Depending on the degree of saturation of the 658 GHz maser we expect the line width of the individual, velocity-blended components across the whole line profile to be approximately equal to the thermal width or well below. Our single dish spectra do not show these spectral components even at the highest spectral resolution. As for VY CMa, 658 GHz imaging is required to identify the individual spectral components. If high saturation dominates, the maser linewidth may broaden up to the thermal width and the maser luminosity may reach an intrinsic maximum value; we then expect a time-stable 658 GHz emission profile and peak intensity. At the other extreme, if unsaturated maser emission dominates, then the exponential light amplification regime dominates and we may expect rapid time variations in the observed 658 GHz spectra. With the present data, we cannot exclude a mixture of saturated and unsaturated maser components toward the same star, which complicates the analysis.  Clear intensity changes would of course strengthen the maser nature of the 658 GHz emission. However, they are difficult to prove and require observations at different epochs using the same observing procedures and calibrators.

To check if there is time variability `hidden' in our 658 GHz spectra, we have compared the ratio of the H$_2$O to $^{13}$CO $J=6-5$ integrated intensities observed with APEX at 658 and  661 GHz with the same intensity ratio observed at a different epoch with Herschel-HIFI. We assume here that the $^{13}$CO broad line profile does not change with the epoch of the observation. This approach is uncertain, however, because it can rub out plausible time variations at specific velocities of the 658 GHz spectrum. Three sources in our sample, o Ceti, IK Tau and W Hya have  been observed by HIFI at a different epoch \citep[][]{justtanont2012}. For these three sources 
we derive from our 2016 observations, 
($\int{{T_A}^{*}}$dv)$_{658}$ / ($\int{{T_A}^{*}}$dv)$_{661}$ = 5.6, 6.1 and 31.4, respectively.
The same ratios taken  from HIFI \citep[][]{justtanont2012} 
are ($\int{T_{mb}}$dv)$_{658}$ / ($\int{T_{mb}}$dv)$_{661}$ = 1.8, 3.5 and 26.3. The APEX and HIFI ratios are relatively similar for W Hya but cannot be reconciled for 
o Ceti and IK Tau even if we include the formal uncertainties which are small for both APEX ($\approx$ 0.8$-$1.5 and 0.1$-$0.5 K.km~s$^{-1}$ at 658 and 661 GHz) and HIFI ($\approx$ 0.1$-$0.3 K.km~s$^{-1}$). This may reflect actual time variation at 658 GHz for o Ceti and IK Tau since we do not expect the thermal $^{13}$CO $J=6-5$ line to vary with time.  However, this result needs to be confirmed with future observations. In addition, we cannot account here for the unknown polarization properties of the 658 GHz maser (see Sect. \ref {sec:relative_intensity}). Comparison of our H$_2$O and $^{13}$CO spectra with the HIFI spectra for VY CMa is perhaps even more uncertain because our APEX data were acquired with two different receiver settings. Nevertheless, the 658 GHz to $^{13}$CO $J=6-5$ integrated intensity ratio obtained with APEX is close to 7.6 
while it is close to 13.5 from the HIFI data \citep[][]{alcolea2013}. 

We further note that the relative 658 GHz peak intensity of W Hya (exhibiting the strongest emission in both \citet[][]{menten1995} and in this work) with respect to VY CMa is close to 1.1 in \citet[][]{menten1995} while it is 3.3 in our data. This may result from time variability in VY CMa which shows an  extraordinary stellar and circumstellar activity, or time variability in both stars at the velocity of their peak intensity. In VY CMa the observations indicate a regular decline of the peak intensity over nearly 25 years. \citet[][]{menten1995}, \citet[][]{hunter2007}, \citet[][]{alcolea2013}, and \citet[][]{richards2014} report peak flux intensities of 2760, 2550, 2026 and 1165 Jy, respectively. Our own data, just taking an average of our Jy to K conversion factor estimates, give 810 Jy. Only images obtained at different epochs could determine which of the circumstellar regions associated with the peak velocity are showing less activity.


\section{Concluding remarks} 
\label{sec:conclusion}

Our main goals were to survey with the APEX telescope a few selected O-rich stars to detect new 658 GHz sources and to investigate some major properties of the 658 GHz transition in the first excited state of the bending mode of water.
Our main findings can be summarized as follows: 

$-$ All sources in our sample of 11 objects (9 AGBs and 2 supergiants) show emission at 658 GHz. This work, together with earlier limited discrete source surveys, suggest that a more extensive survey of evolved stars should discover many more objects and confirm that the 658 GHz emission is widespread. (We are currently conducting such a survey with the APEX telescope.)

$-$ We have given simple arguments, supported by modeling, to show that the 658 GHz emission is masing. We suspect that some time variability is embedded in our data but we do not exclude that the 658 GHz line profile is composed of velocity-blended saturated maser components. 

$-$ We have found a loose correlation of the 658 GHz velocity extent with the SiO, v=1, $J=2-1$ velocity extent which still needs to be confirmed with more observational data. This correlation suggests that the physical conditions required to excite both masers and give long enough emission paths to be detectable are similar. Hence, both masers are excited in regions close to the central star. 

$-$ The simple line shape and sometimes strong  658 GHz emission make the 658 GHz sources potentially good ALMA Band 9 phase calibrators. The most interesting objects deserve to be imaged with ALMA to unravel the spatial structure and the exact nature of the 658 GHz emission.

 { \it Acknowledgements.} This publication is based on data acquired with the Atacama Pathfinder Experiment (APEX). APEX is a collaboration between the Max-Planck-Institut fur Radioastronomie, the European Southern Observatory, and the Onsala Space Observatory. We thank APEX staff for carrying out these observations. We also thank the referee for useful comments.

\bibliographystyle{aa}
\bibliography{biblio-copie}
\end{document}